\documentclass[conference,10pt]{IEEEtran}

\usepackage{comment}
\usepackage{epsfig}
\usepackage{amssymb}
\usepackage{cite}
\ifCLASSINFOpdf
\else
\fi
\usepackage[cmex10]{amsmath}
\usepackage{amsfonts}

\usepackage{algorithm}
\usepackage{algorithmic}

\usepackage{mdwmath}
\usepackage{mdwtab}
\usepackage[tight,footnotesize]{subfigure}
\usepackage{float}
\usepackage{subfigure}
\usepackage{booktabs}

\newtheorem{lemma0}{\bf Lemma}

\newtheorem{definition0}{\bf Definition}

\begin{document}
%
\title{Behavior-Based Online Incentive Mechanism for Crowd Sensing with Budget Constraints}
\author{\IEEEauthorblockN{Jiajun Sun and
Huadong Ma
}
\IEEEauthorblockA{
Beijing University of Posts and Telecommunications, 
Beijing 100876, China\\
Email: jiajunsun.bupt@gmail.com}
}


\maketitle

\begin{abstract}
Crowd sensing is a new paradigm which leverages the ubiquity of
sensor-equipped mobile devices to collect data. To achieve good
quality for crowd sensing, incentive mechanisms are indispensable to
attract more participants. Most of existing mechanisms focus on the
expected utility prior to sensing, ignoring the risk of low quality
solution and privacy leakage. Traditional incentive mechanisms such
as the Vickrey-Clarke-Groves (VCG) mechanism and its variants are
not applicable here. In this paper, to address these challenges, we
propose a behavior based incentive mechanism for crowd sensing
applications with budget constraints by applying sequential all-pay
auctions in mobile social networks (MSNs), not only to consider the
effects of extensive user participation, but also to maximize high
quality of sensing data submission for the platform (crowd sensing
organizer) under the budget constraints, where users arrive in a
sequential order. Through an extensive simulation, results indicate
that incentive mechanisms in our proposed framework outperform
existing solutions.
\end{abstract}

%
\IEEEpeerreviewmaketitle

\section{Introduction}\label{sec:intro}
With the increasing ubiquity of
sensor-embedded
mobile devices (e.g., smartphones), 
mobile social networks (MSNs), which integrate data collection
techniques and services into many kinds of social networks
\cite{aggarwal2011integrating,fan2013geocommunity}, have received
considerable research efforts in recent years due to two changes as
follows. First, the terminal devices for social network applications
change from PCs to mobile phones. Second, the interactive mode
extends from the virtual space to the real physical world. MSNs
provide a new opportunity for crowd sensing, which takes advantage
of the pervasive mobile devices to solve complex sensing tasks. A
typical example of crowd sensing applications is to provide the
support for green traffic by sensing and reporting timely the
measurements about
traffic flows in some region. 
Different from existing crowdsourcing systems, crowd sensing
exploits sensing and processing abilities of mobile devices to
provide sensing data from the real physical world towards a specific
goal or as part of a social or technical experiment
\cite{ra2012medusa,zhou2013consub}.


Extensive user participation and submission quality are two crucial
factors determining whether crowd sensing applications in MSNs can
achieve good service quality. Most of the current crowd sensing
applications are based on a common hypothesis that all users
voluntarily participate in submitting the sensing data. However,
mobile devices are controlled by rational users, in order to
conserve energy, storage and computing resources, so selfish users
could be reluctant to participate in sensing data for crowd sensing
applications. Thus, it is indispensable to provide some incentive
schemes to stimulate selfish participants to cooperate in MSNs. Only
a handful of works
\cite{yang2012crowdsourcing,lee2010sell,duan2012incentive,jaimes2012location}
focus on incentive mechanism for crowd sensing applications. All of
these works apply a regular auction (e.g., a reverse auction) only
for off-line crowd sensing applications with the ex-ante payment
commonly known as ``Free rider problem".

The data submission quality issue from participants is also
challenging in crowd sensing applications
\cite{liu2011crowdsourcing}. If the submission quality of
participants is not well guaranteed, although the extensive user
participation offers useful information, the service quality from
participants is far from satisfactory to the requesters of crowd
sensing applications. For example, on the one hand, the limit of the
coverage constraint may make the participants with high quality data
drop out of crowd sensing application \cite{jaimes2012location}; On
the other hand, traditional incentive mechanisms such as the
Vickrey-Clarke-Groves (VCG) mechanism \footnote{Vickrey auction is a
type of sealed-bid auction, where bidders submit written bids
without knowing the bid of the other people in the auction, and in
which the highest bidder wins, ``but the price paid is the
second-highest bid.''} and its variants also will make the
participants with higher true valuation become starved frequently to
win, thereby drop out of crowd sensing applications
\cite{lee2010dynamic}. Therefore, special mechanisms must be
included to handle these challenges.

Although both extensive user participation and submission quality
issues have been identified as two crucial human factors for crowd
sensing applications, many recent research works
\cite{yang2012crowdsourcing,lee2010sell,duan2012incentive,jaimes2012location,liu2011crowdsourcing,lee2010dynamic,ghosh2011incentivizing,ren2012maximizing}
tend to separately study them in crowd sensing applications. The
reason is that, if the extensive user participation and submission
quality problems are addressed at the same time for crowd sensing
applications, the issue would become more challenging. For example,
some submission quality enhanced techniques
\cite{ghosh2011incentivizing,ren2012maximizing} stimulate
participants to generate high quality sensing contents to achieve
good service quality, but they could make some incentive strategies,
especially the reputation-based incentive strategies under budget
constraints, hard to implement extensive user participation coverage
constraints for crowd sensing applications, since it is not
practical to assume that the requester will always provide an
unlimited budget to achieve good service quality. Therefore, how to
simultaneously address both extensive user participation and
submission quality issues becomes particularly challenging for crowd
sensing applications with budget constraints.


In this paper, to address the fore-mentioned challenges, we propose
a behavior based incentive mechanism for practical crowd sensing
applications with budget constraints. Specifically speaking, our
main contributions are summarized as follows:
\begin{itemize}
\item We explore a
behavior based incentive mechanism for crowd sensing applications
with budget constraints in MSNs. In order to simultaneously satisfy
the requirements of both extensive user participation and high
quality sensing data submission, we combine the all-pay auction
theory and a proportional share allocation rule to stimulate the
participants to generate high efforts and adequate coverage
constraints to achieve the better service for the requester of crowd
sensing applications with
budget constraints. 
\item We focus on a more real crowd sensing scenario where users arrive one by one online in a random order.
We model the issue as a sequential all-pay auction \footnote{All-pay
auction, is an auction in which every bidder must pay regardless of
whether he wins the prize, which is awarded to the highest bidder as
in a conventional auction.}, in which sensing data are submitted
sequentially and the users with the high quality sensing data are
selected as the winners. Further, after observing previous
submissions, we derive every user best response effort bidding
function for sequential crowd sensing applications with budget
constraints, which influences user participation and sensing data
submission quality.
\item Extensive simulations
show that our proposed incentive mechanism outperforms the existing
solutions.
\end{itemize}
%

The rest of the paper is organized as follows. In Section
\ref{related}, we briefly discuss the related work and motivation.
In Section \ref{SystemModel}, we present our system model, related
definitions and our design goals. In Section \ref{MecahnismDesgn},
we design a behavior based incentive strategy for sequential crowd
sensing in MSNs, and present the performance evaluation in Section
\ref{Experiment}. Finally, Section \ref{Conclude} concludes the
paper.

\section{Background and Related Work}~\label{related}
Extensive user participation and submission quality issues are two
crucial human factors for crowd sensing applications in MSNs. The
authors of \cite{reddy2010recruitment} proposed recruitment
frameworks to enable the platform to identify well-suited
participants for sensing data collections. However, they only
considered the users' selection, rather than the incentive mechanism
design. In recent years, most of reported studies have focused on
how to stimulate selfish participants to enhance user participation
levels. For instance, the authors of
\cite{duan2012incentive,lee2010dynamic,jaimes2012location,sun13globecom}
explored the extensive user participation to achieve a good sensing
service for crowd sensing applications. Obviously, it is not
practical to assume that the requester in their mechanisms will
always have an unlimited budget. The authors of
\cite{chen2011approximability,singer2010budget,tran2012efficient,yang2012crowdsourcing}
designed an incentive mechanism to enhance user participation levels
under a budget constraint. Although they designed truthful
mechanisms, which optimized the utility function of the platform
under a fixed budget constraint,
to incentive extensive user participation, the effects of the online sequential manner, in which users arrive, were neglected. 
In practice, recently, there are a few works focusing on both budget
constraints and the online sequential manner of users' arrival to
enhance user participating levels. For instance, the authors of
\cite{badanidiyuru2012learning,sun14ICCcollection} exploited posted
price mechanisms for stimulating the online arrival user
participation. The authors of \cite{singer2013pricing} leveraged
threshold price mechanism for maximizing the number of tasks under
budget constraints and task completion deadlines. However, they did
not consider the submission quality issue of sensing data.

Compared with the extensive user participation issue, there are only
a handful of research works
\cite{ghosh2011incentivizing,ren2012maximizing} focusing on the
submission quality issue for crowd sensing applications in MSNs.
These works stimulate participants to submit high quality sensing
data to achieve good service quality, but do not support the
extensive user participation issue. The authors of
\cite{pham2011novel} study a problem about the low  payment from  a
predefined number of participants. On the contrary, we focus on the
utility maximum problem with a given budget. In our mechanism, in
order to simultaneously satisfy the requirements of both extensive
user participation and high quality sensing data submission, we
combine the all-pay auction mechanism and a proportional share
allocation rule to achieve the better service for crowd sensing
applications with budget constraints. Furthermore, we account for
the online arrival of users and model the issue as an online
sequential all-pay auction. Simulations indicate that our proposed
incentive mechanisms outperform existing solutions.

\section{System Model and Problem Formulation}~\label{SystemModel}
\subsection{System Model}
\begin{figure}
\setlength{\abovecaptionskip}{0pt}
\setlength{\belowcaptionskip}{10pt} \centering
\centering
\includegraphics[width=0.25\textwidth]{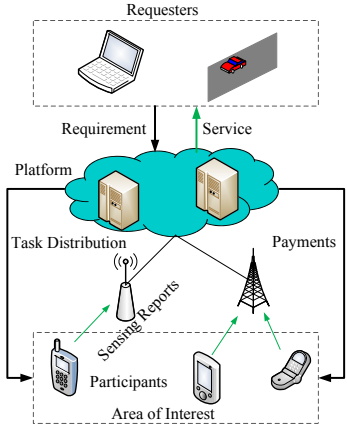}
\caption{Our crowd sensing system framework with all-pay auctions.}
\label{crowd}
\vspace{-15pt}
\end{figure}

We consider the following crowd sensing system model illustrated in
Fig.~\ref{crowd}. The system consists of a crowd sensing application
platform, to which a requester with a budget $B>0$ posts a crowd
sensing application that resides in the cloud and consists of
multiple sensing servers, and many mobile device users, which are
connected to the cloud by cellular networks (e.g., GSM/3G/4G) or
WiFi connections. The crowd sensing application first publicizes a
sequential sensing task in an Area of Interest (AoI), denoted by
$\Gamma=\{\tau_{1},\tau_{2},\cdots, \tau_{m}\}$. Assume that a set
of users $\mathcal {U}=\{1,2,\cdots,n\}$ interested in the crowd
sensing service campaign register to the platform, and then arrive
online in a sequential order.

\subsection{User Model}
We explore online auctions where each user arriving randomly has a
chance to win the auction, and each offering sensing data such as
location information for constructing network coverage maps, for
which each user associate a private cost $c_{i}$. The platform has a
public (known to the mechanism designer and maybe known to all
agents) budget $B\in\mathbb{R}_{+}$, and a public nondecreasing
utility function $U$ over the subsets of users. We assume that in
each time step, a single user appears and the platform makes a
decision that is based on the information it has about the user and
the history of the previous $i-1$ stages. Generally, there are three
classes of user models: \textit{the i.i.d. model}, \textit{the
secretary model}, and \textit{the adversarial model}. The first
model means that at each time step the costs and values of users are
drawn from some unknown distributions. The second model means that
the users' costs are chosen by an adversary, however their arrival
order is a permutation that is drawn uniformly at random from the
set of all possible permutations. In the third model, the users'
costs and their arrival order are chosen by an adversary. Note that
in the third model, although the adversary cannot observe the
actions the mechanism takes, since it has full knowledge, the
adversary chooses the worst arrival order and costs. Thereby, the
mechanism cannot obtain the optimal solutions. Thus, in this paper,
we only account for the two models with respect to the distribution
of users, described in increasing order of generality: the
i.i.d.model and the secretary model.

\subsection{Problem Formulation}
We model the above interactive process between the platform and
users as an online sequential all-pay auction with the budget
constraint, where the plethora of users with different preference
and skill ability $\theta \in[\underline{\theta},
\overline{\theta}]$ ($\underline{\theta}$ and $\overline{\theta}$
denote the least skilled behavior ability and the most skilled
behavior ability respectively) arrive and compete in a contest by
their efforts in one period. User behavior abilities are described
by the distribution function $F$. Assume that users are
game-theoretic and seek to make strategy to maximize their
individual utility in equilibrium. Receiving the crowd sensing
campaign from the platform, each user $i$ provides its sensing
profile $\Gamma_{i}$ to the platform according to its preference and
skill ability so as to expect a prize allocation (i.e.,payment) in
return for its sensing data submission. The platform determines the
payment $M_{1}, M_{2}, \cdots, M_{n}$ to the top $n$
(highest-ranked) submissions in one period $(M_{1}\geq M_{2}
\geq\cdots \geq M_{L}=m>M_{L+1}=\cdots=M_{n}=0,
\sum_{l=1}^{n}M_{l}=B)$: the user with the best submission receives
$M_{1}$, the first runner-up receives $M_{2}$, and so forth. Note
that different from traditional mechanisms, here $\Gamma_{i}$ is not
fixed. Thereby, the contribution of user $i$, denoted by $e_{i}$, is
uncertain and depends on user's efforts. But when these sensing data
are submitted to the platform, the platform can identify their
contributions, i.e., their marginal utility (to be elaborated
later).

More formally, an online mechanism $\mathcal{M}=(f,p)$, which
consists of an allocation function $f:
\mathcal{R}_{+}^{n}\rightarrow 2^{[n]}$ and a payment function $p:
\mathcal{R}_{+}^{n}\rightarrow\mathcal{R}_{+}^{n}$, is needed. That
is, for any random arriving users'
$\hat{\mathcal{P}}=(\Gamma_{1},\Gamma_{2},\cdots,\Gamma_{n})$
strategy sequence, the allocation function computes an allocation of
the budget for participatory users $S^{'}\subseteq \mathcal{U}$ and
the payment function returns a payment vector to participatory
users. Thus, the utility of user $i$ is
$p_{i}-\frac{e_{i}}{\theta_{i}}$ if it is allocated,
$-\frac{e_{i}}{\theta_{i}}$ otherwise. The platform expect to obtain
the maximum value from the particpatory users' service quality under
the budget constraint, i.e.,

\begin{equation*}
\max U(S^{'}) ~~~~Subject ~to ~\sum_{i\in S^{'}} p_{i}\leq B
\end{equation*}
where $U(S^{'})$ is the monotone submodular value function of
services from the participatory users $S^{'}$. More variables are
given in Table \ref{tab:notations}.

\begin{table}[t]
\begin{center}\caption{Summary of Notations} \label{tab:notations}
\begin{tabular}{cc}\toprule
Variable & Description\\ \midrule
$\mathcal {U}$, $n$, $i$ & set of users, number of users, and some user\\
$B, B^{'}, e^{*}$ & budget constraint and stage-budget, the effort threshold\\
$T, T^{'}, t$ & deadline, end time step of each stage, and each time step\\
$\Gamma_{i}$ & user $i$' sensing profile\\
$\Gamma, m, \tau_{j}$ & set of services, number of services, and some service\\
$\theta_{i}, e_{i}, U_{i}$ & behavior ability, effort bid and marginal utility of user $i$  \\
$S^{'}, \mathcal{J}$ & set of sampled users and winners in sampled users\\
$L$ & number of users with allocated prizes larger than zero\\
$p_{i}, F_i$ & number of allocated prizes and ability distribution of user $i$\\
$\underline{\theta},\overline{\theta}$ & the least skilled and the most skilled behavior ability\\
$\Phi_{j}$ & order cumulative distribution function of the $j$-th best type \\
 \bottomrule
\end{tabular}
\end{center}
\vspace{-20pt}
\end{table}

Putting the above online policy together, we design incentive
mechanisms based on users' behavior abilities to simultaneously
satisfy the requirements of both extensive user participation and
high quality sensing data submission. To stimulate users to produce
endogenous maximal efforts, we introduce a sequential all-pay
auction to the above online allocation strategies for our incentive
mechanism design. Considering a risk-neutral budget constrained
all-pay auctions, the above utility function $U(S^{'})$ indicates
that $U(e_{1},\cdots,e_{L})=\sum _{i\in S^{'}}f(e_{i})$, where the
effort $f(e_{i})$ is equal to the marginal utility of user $i$,
i.e., $f(e_{i})=U_{i}(S^{'})=U(S^{'}\cup\{i\})-U(S^{'})$ when the
set of sampled users is $S^{'}$.

Generally speaking, the goal of the platform is to select a subset
$S$ with the size $L$
that maximizes the total efforts of the users under the given budget. 
A utility-maximizing platform should select the number of prizes
$L$, the total prize budget B and the allocation of prizes
$M_{1}+M_{2}+\cdots+M_{L}=B$ which maximizes the platform's utility
$U(e_{1},\cdots,e_{L},B)$. More generally, given a budget $B$ and a
reserved effort $m$, finding a subset $S$, is equivalent to
maximizing the above coverage issue of sensing data submitted by
users.

\textbf{A case study:} We consider the following scenario where the
platform expects to obtain the sensing data covering all roads in an
AoI. For convenience of presentations and calculations, we divide
each road in the AoI into multiple discrete points, and the
objective of the platform is equivalent to receiving the submitted
sensing data covering all points before $T$. The set of all points
in the AOI is denoted by $\Gamma=\{\tau_{1},\tau_{2},\cdots,
\tau_{m}\}$. The set of points submitted by user $i$ is denoted by
$\Gamma_{i}$ (called user $i$'s sensing profile), which are the
result of users' efforts based on their behavior ability. For ease
of presentation, assume that each point $\tau_{j}$ needs to be
sensed at most one time. The value of the sampled users to the
platform is: $U(S^{'})=\sum\limits_{j = 1}^m {\min \left\{
{1,\sum\nolimits_{i \in S^{'}} {s_{i,j} } } \right\}}$, where
$s_{i,j}$ equals to $1$ if $\tau_{i}\in\Gamma_{i}$, $0$ otherwise.

\section{Optimal Mechanism Design for The Sensing Contest} \label{MecahnismDesgn}
In this section, we present an online contest mechanism for
achieving extensive user participation and stimulating users to
submit high quality sensing data. To this end, we adopt a
multiple-stage sampling-accepting process for guaranteeing extensive
user participation. Then we introduce an online sequential all-pay
auction for stimulating users to submit high quality sensing data.
\subsection{Incentive Mechanism Design}
An online contest mechanism needs to overcome several nontrivial
challenges. First, the users' effort bids are unknown and need to be
elicited in a truthful reporting manner. Second, the total prizes
cannot exceed the platform's budget; Thirdly, the mechanism needs to
tackle the online arrival of the users and  makes irrevocable
decisions on whether to allocate the number of prizes; Finally, and
most important, the mechanism needs to cope with strategic users'
endogenous efforts. To achieve good service quality, previous online
solutions and generalized secretary problems
\cite{hajiaghayi2004adaptive,bateni2010submodular,singer2013pricing,kleinberg2005multiple}
is via sampling: the first batch of the input is rejected and used
as a sample which enables making an informed decision on the rest of
the users. However, since users are likely to be discouraged to
sense data knowing the pricing mechanism will automatically reject
their effort bid. In other words, those users arriving early have no
incentive to report their bids to the platform, which may delay the
users' completion or even lead to task starvation. 
On the other hand, the above mechanisms only apply for fixed
services submitted by each user, thereby neglect users' endogenous
efforts.

To address the above challenges, we adopt a multiple-stage
sampling-allocating process to design our online contest incentive
mechanism. At each stage, based on the above submodularity, the
mechanism maintains a effort threshold which is used to decide to
allocate the number of the users' prizes. The mechanism dynamically
increases the sample size and learns a budget that are enough to
allocate users for maximizing the total utility, then apply this
density threshold for making further allocation decisions.

\begin{figure}
\setlength{\abovecaptionskip}{0pt}
\setlength{\belowcaptionskip}{10pt} \centering
\centering
\includegraphics[width=0.42\textwidth,height=0.12\textwidth]{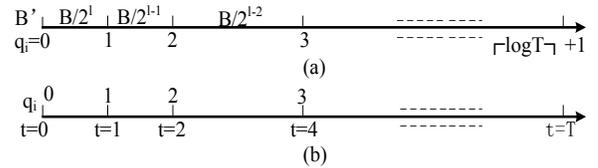}
\caption{Illustration of a multi-stage sample process with deadlines
$T$. (a)Budget $B^{'}$ versus quantile $q_{i}$; (b)Quantile $q_{i}$
versus time $t$.} \label{multistage}
\vspace{-15pt}
\end{figure}

Specifically, given a distribution on the arrival of users, we can
easily calculate every time step $t$ s.t. the probability that a
user arrives before t is $1/2^{i}$. All of $T$ time steps are
divided into $2^{i}$ quantiles: $\{0,1,\cdots,\lceil\log T\rceil\}$.
We apply $S^{'}$ to denote the set of all sampled users until the
time step $t$. Our mechanism (see Algorithm \ref{prizestructure})
iterates over $q_{i}\in\{0,1,\cdots,\lceil\log T\rceil\}$ and at
every time step $q_{i}$, a budget of $B/2^{i}$ is applied to
allocate the number of prizes (illustrated in
Fig.~\ref{multistage}). Firstly, when a new user $i$ arrives, with
probability $1/3$, our mechanism computes the best response effort
bid of user $i$ according to his current behavior ability, and
thereby can submit its sensing data as a result of his efforts.
Receiving user $i$'s sensing data $\Gamma_{i}$, regardless of
whether it is given prizes, it is added to the sample set $S^{'}$
due to the nature of the all-pay auction. Given a set of sampled
users $S^{'}$, the platform computes the marginal utility of user
$i$, i.e., $U_{i}(S^{'})=U(S^{'}\cup\{i\})-U(S^{'})$, $i\notin
S^{'}$. The mechanism allocates prizes to user $i$ as long as the
minimal prizes of the last offline sample stage is not larger than
the result of effort threshold $e$ (to be elaborated in the
following subsection) multiplied by the marginal utility $U_{i}(S)$,
and the allocated stage-budget $B'$ has not been exhausted. At the
first stage, we set a small effort threshold and minimal prizes to
start our mechanism. The randomization addresses extreme cases in
which only a single user with the strongest behavior abilities can
complete a large fraction of the crowd sensing application at the
threshold efforts. This is the result that an incentive compatible
variant of Dynkin's celebrated algorithm \cite{dynkin1963optimum} to
the issue of hiring the best secretary, is tailored to our setting.

To make the above mechanism to achieve our goal, in the following,
we elaborate the computation methods of the effort threshold  and
the arrival user's optimal sequential efforts.

\begin{algorithm}[t] 
\renewcommand{\algorithmicrequire}{\textbf{Input:}}
\renewcommand\algorithmicensure {\textbf{Output:} }
\caption{The Budgeted Behavior-based incentive mechanism under
Sequential all-pay auctions(BBS)} 
\label{prizestructure} 
\begin{algorithmic}[1] 
\REQUIRE Budget constraint $R$, deadlines $T$ \\
\STATE
$(T^{'},B^{'},S^{'},e^{*},M^{*},t)\leftarrow(\frac{T}{2^{\lfloor\log_{2} T\rfloor}},\frac{B}{2^{\lfloor\log_{2} T\rfloor}},\emptyset,\varepsilon,\nu,1)$;\\

\FOR{$t\leq T$}
            \IF{there is a user $i$ arriving at time step $t$}\label{threeline1}
            \STATE With probability 1/3 do:\\
                \STATE Exert the effort $e_{i}$ for user $i$ obtained according to his behavior abilities by using the expression (\ref{bideffort})
                and submit its sensing data $\Gamma_{i}$ as its effort result to the platform;\\
                \STATE Receiving these data $\Gamma_{i}$, the platform computes the marginal utility of user $i$, i.e., $U_{i}(S^{'})$;\\
                \IF {$M^{*}\leq e^{*}\cdot U_{i}(S^{'})\leq B^{'}-\sum_{j\in S}{p_{j}}$}\label{fourline1}
                    \STATE $p_{i}\leftarrow e\cdot U_{i}(S^{'})$;\label{fiveline1}\\
                \ENDIF
                \STATE $S^{'}\leftarrow S^{'}\cup\{i\}$;\\
            \STATE Otherwise do:
                   \FOR {the first user $i$ arriving by $\lfloor T^{'}\rfloor$}
                       \STATE  Run Dynkin's algorithm and offer $B$ the winner; \\
                   \ENDFOR

            \ENDIF\label{eightline1}
            \IF{$t=\lfloor T^{'}\rfloor$}
                \STATE Calculate $(e^{*}, M^{*})$=GetEffortThreshold($S^{'}$, 2$B^{'}$);\\
                \STATE set $B^{'}\leftarrow 2B^{'}$, $T^{'}\leftarrow 2T^{'}$;\\
            \ENDIF
            \STATE $t\leftarrow t+1$;\label{thirteenline1}\\
\ENDFOR
\end{algorithmic}
\end{algorithm}

\subsection{Threshold Effort Decision}
In this subsection, we first introduce a threshold effort to ensure
the extensive participation. Then we apply all-pay auction to
enhance users' submission quality. We now turn to the following
definition of nondecreasing submodular functions used in our pricing
mechanisms.

\begin{definition0}[\textbf{Submodular Function}]\label{df:nashequilibrium}
Let $\mathbb{N}$ be a finite set, a function $U$ : $2^{\mathcal
{U}}\rightarrow \mathbb{R}$ (the set of reals) is submodular if
$U(S\cup\{i\})-U(S)\geq U(S^{'}\cup\{i\})-U(S^{'}), \forall
S\subseteq S^{'}\subseteq \mathcal {U}$, where $S, S^{'}$, and
$\mathcal {U}$ are illustrated in Table \ref{tab:notations}, and $i$
denotes some user.
\end{definition0}

\indent\textbf{Background:} Submodularity, a discrete analog of
convexity, has played an essential role in combinatorial
optimization \cite{lovasz1983submodular}. It appears in many
important settings and almost everywhere
\cite{goemans2009approximating} including cuts in graphs
\cite{goemans1995improved,queyranne1995combinatorial,iwata2001combinatorial},
rank function of matroids \cite{edmonds2003submodular}, set covering
problems and plant location problems \cite{feige1998threshold}. In
many settings such as set covering or matroid optimization, the
relevant submodular functions are monotone, meaning that $U(S)\leq
U(S^{'})$ whenever $S\subseteq S^{'}$. More recently submodular
functions have become key concepts both in the machine learning and
algorithmic game theory communities. For example, submodular
functions have been used to model bidders' valuation functions in
combinatorial auctions \cite{lehmann2001combinatorial,
dobzinski2006truthful, balcan2011learning}, and for solving feature
selection problems in graphical models \cite{krause2012near} or for
solving various clustering problems \cite{narasimhan2007local}.

\begin{lemma0}\label{expectedsubmodular}
The value function $U(S)$ is monotone submodular.
\end{lemma0}

The proof of Lemma \ref{expectedsubmodular} is given in the
Appendix.



In a general submodular maximization problem, a proportional share
allocation rule is a natural fit to compute the effort threshold
(see Algorithm \ref{efforttheshold}) due to its monotonicity when
users are sorted according to their efforts relative to increasing
marginal contributions. 
%
However, to enhance the sensing data quality (users' efforts), we
apply the optimal winning participant number $L$ and the optimal
prize amounts $M_{i} (i\in\{1,\cdots, L\})$ to calculate the effort
threshold of the next time step. The optimal winning participant
number $L$ and the optimal prize amounts $M_{i} (i\in\{1,\cdots,
L\})$ can be calculated according to the method in
\cite{archak2009optimal}.
Then, these sampled users are sorted according to increasing
marginal contributions relative to their prize amounts. This sorting
implies:
\begin{equation*}
U_{1}/ M_{1}\geq U_{2}/ M_{2}\geq \cdots \geq U_{L}/ M_{L},
\end{equation*}
where $U_{i}$ denotes $U_{i\mid S^{'}_{i-1}}=(U(
S^{'}_{i-1}\cup\{i\})-U(S^{'}_{i-1})$), $ S^{'}_{i-1}=\{1,2,
\cdots,i-1\}$, and $S^{'}_{0}=\emptyset$. The specifical iteration
process is illustrated in Algorithm \ref{efforttheshold} to
guarantee the extensive user participation.

\begin{algorithm}[t] 
\renewcommand{\algorithmicrequire}{\textbf{Input:}}
\renewcommand\algorithmicensure {\textbf{Output:} }
\caption{GetEffortThreshold} 
\label{efforttheshold} 
\begin{algorithmic}[1] 
\REQUIRE Sample set $S^{'}$, stage budget constraint
$B^{'}$ \\
\ENSURE Effort threshold and minimal prizes, i.e., $(e, M_{L})$. \\ 
\STATE Compute the optimal winners' number and the optimal prize
amounts $M_{i} (i\in\{1,\cdots, L\})$;\\
\STATE Sort their marginal utility relative to their prize amounts,
s.t. $U_{1}/ M_{1}\geq U_{2}/ M_{2}\geq \cdots \geq U_{L}/ M_{L}$; set $k=1$;\\

\STATE The winner set $\mathcal {J}\leftarrow \emptyset$;
$i\leftarrow\arg\max_{j\in\mathcal {X^{'}}}(U_{j}(\mathcal
{J})/M_{j})$;\\
            \WHILE{$k\leq L$ and $M_{i}\leq U_{i}B^{'}/U(\mathcal {J}\cup i)$}
                    \STATE $\mathcal {J}\leftarrow \mathcal {J}\cup i$;\\
                    \STATE $i\leftarrow\arg\max_{j\in S^{'}\setminus\mathcal {J}}(U_{j}(\mathcal{J})/M_{j})$;\\
                    \STATE $k\leftarrow k+1$;\\
            \ENDWHILE
            \STATE $e \leftarrow \mathcal{B^{'}}/U(\mathcal {J})$;\\
            \RETURN $(e, M_{L})$;\\
\end{algorithmic}
\end{algorithm}
\subsection{Computing Sequential Efforts}
In the following subsection, due to space limitations, assuming that
the reserved value $m$ is zero, i.e., $e_{0}=0$, we derive the
equilibrium effort bidding function for sequential all-pay auction
arrival users. For the positive reserve case, $e_{0}>0$, the similar
results can also be derived, which will be discussed in our future
work.

For technical reasons, we assume that ties are broken in favor of
the late entrant \footnote{This is a technical assumption to derive
strict subgame perfect equilibria instead of $\epsilon$-equilibria.}
so as to illicit the truthful effort bid of the current user. For
user $i$, a submission of quality $e_{i}$ costs $e_{i}/\theta_{i}$,
indicating that it is less costly for a high ability user to submit
a sensing data of a given quality than a low ability user. Assume
that $\Phi_j$ is the cumulative distribution function of the $j$-th
best type out of $n-1$ users. According to equation (2.1.3) of
\cite{david2003order}, we have
\begin{equation*}
\Phi_j(e)=\sum\limits_{i = 1}^{n - 1} {\left( {\begin{array}{*{20}c}
   {n - 1}  \\
   i  \\
\end{array}} \right)} \Phi (e)^i (1 - \Phi (e))^{n - 1 - i}.
\end{equation*}

Moreover, the expected utility $\overline V$ of the user of ability
type $\theta$ bidding with quality $e$ is $\overline
V=\sum\limits_{l =
1}^L {\Phi _j (e)(M_l-M_{l + 1})}$. 
Thus, applying backward induction iteratively for user $n,
n-1,\cdots,i+1$, we can obtain the following maximization problem of
the just arrival user $i$:

\begin{equation}\label{effortbid}
\begin{array}{l}
 \mathop {\max }\limits_{e_i } \{ \overline V \prod\limits_{j = i + 1}^n {F _j (e_j  = 0)}  - \frac{{e_i }}{{\theta _i }}\} \\
 s.t.~~~e_i  \ge  e_{L-th} (\theta _{L-th}), \\
 \end{array}
\end{equation}
where $e_{L-th} (\theta _{L-th})$ denotes the L-th largest effort
bids observed by user $i$.

Therefore, in the following, we derive user $i$'s best response
effort bidding function based on the assumption that the behavior
ability distribution function is $F(x)=x^{c}$, where ${0<c<1}$ like
\cite{sela2012multi,liu2011crowdsourcing}. To derive a closed-form
solution, we therefore restrict attention to the specific family of
concave distribution functions, for which we are able to explicitly
calculate the subgame perfect equilibrium e¤ort of each contestant.
This will allow us to derive some general optimal results on
multi-stage sequential all pay auctions. If we let ${\overleftarrow
{\theta _i } =[(1 - d_{L - th} )\theta _{L - th} ]^{d_i /d_{L - th}
} }$, and ${\overrightarrow {\theta _i }= [(1 - d_{1 - th} )\theta
_{1 - th} ]^{d_i /d_{1 - th} } /c}$, given
$e_{1},e_{2},\cdots,e_{i-1}$, user $i$'s best response effort
bidding function is obtained as follows (the complete calculation
process given in the Appendix):

\begin{equation}\label{bideffort}
e_i  = \left\{ {\begin{array}{*{20}c}
   {\begin{array}{*{10}c}
   {0~~~~~~~~~~~~~~~~~~~~~~~~~~~~~~~}  & {0 \le \theta _i  < \overleftarrow {\theta _i } }  \\
\end{array}}  \\
   {\begin{array}{*{20}c}
   {\{ e_{j - th} (\theta _{j - th} )\} _{j \in \{ 1,2, \cdots ,L\} } }  & {\overleftarrow {\theta _i }  \le \theta _i  < \overrightarrow {\theta _i } }  \\
\end{array}}  \\
   {\begin{array}{*{20}c}
   {\overline V [\theta _i (1 - d_i )]^{1/d_i } ~~~~~~~~~~~~}  & {\overrightarrow {\theta _i }  \le \theta _i  \le 1}  \\
\end{array}}  \\
\end{array}} \right.
\end{equation}

\section{Performance Evaluation}~\label{Experiment}
To evaluate the performance of our BBS mechanism, and explore the
effects of extensive user participation and high quality sensing
data submission for real crowd sensing applications, we implement
the mechanism by applying the well-known Manhattan model obtained
from the Google Map, which is the same as \cite{sheng2012energy}.

\subsection{Experimental Setup}

In the simulation, the sensing range of each mobile phone is set to
7 meters. The AoI obtained from the Google Map is located at
Manhattan, NY, which spans $4$ blocks from west to east with a total
length of $1.135km$ and $4$ blocks from south to north with a total
width of $0.319 km$, and includes 3 Avenues and 3 Streets. We divide
each road in the AoI into multiple discrete AoIs with a uniform
spacing of 1 meter, so the AoI consists of $1135*3+319*3-9=4353$
PoIs in total, where each Avenue has $1135$ points and each Street
has $319$ points, and the crossing of Avenues and Streets includes
$9$ points. Data sensing area sizes are set at random. In summary,
we used $1887$ sensing data packets from different locations via
$200$ different users \cite{sheng2012energy}. The sensing range
(behavior ability) of different users' sensor changes from $3$
meters to $10$ meters. To start the mechanism, we initially set the
effort threshold $e^{*}$ and the number of minimal prizes $M^{*}$ of
Algorithm \ref{prizestructure} are set to $0.1$. Users arrive
according to a Poisson process in time with arrival rate $\lambda$,
which varies from 0.3 to 8 with the increment of 0.1. Arriving at
our scenario, some user is placed randomly at some location on the
AoI's roads. All measurements are averaged over $30$ sensing tasks.
Our primary goals are to evaluate the performance of the online
mechanism on real effort bids as well as to test users'
participating response to different mechanisms.


\begin{figure*}
\vspace{10pt} \centering \subfigure[]{ \label{threshold}
\includegraphics[width=1.7in, height=1.3in]{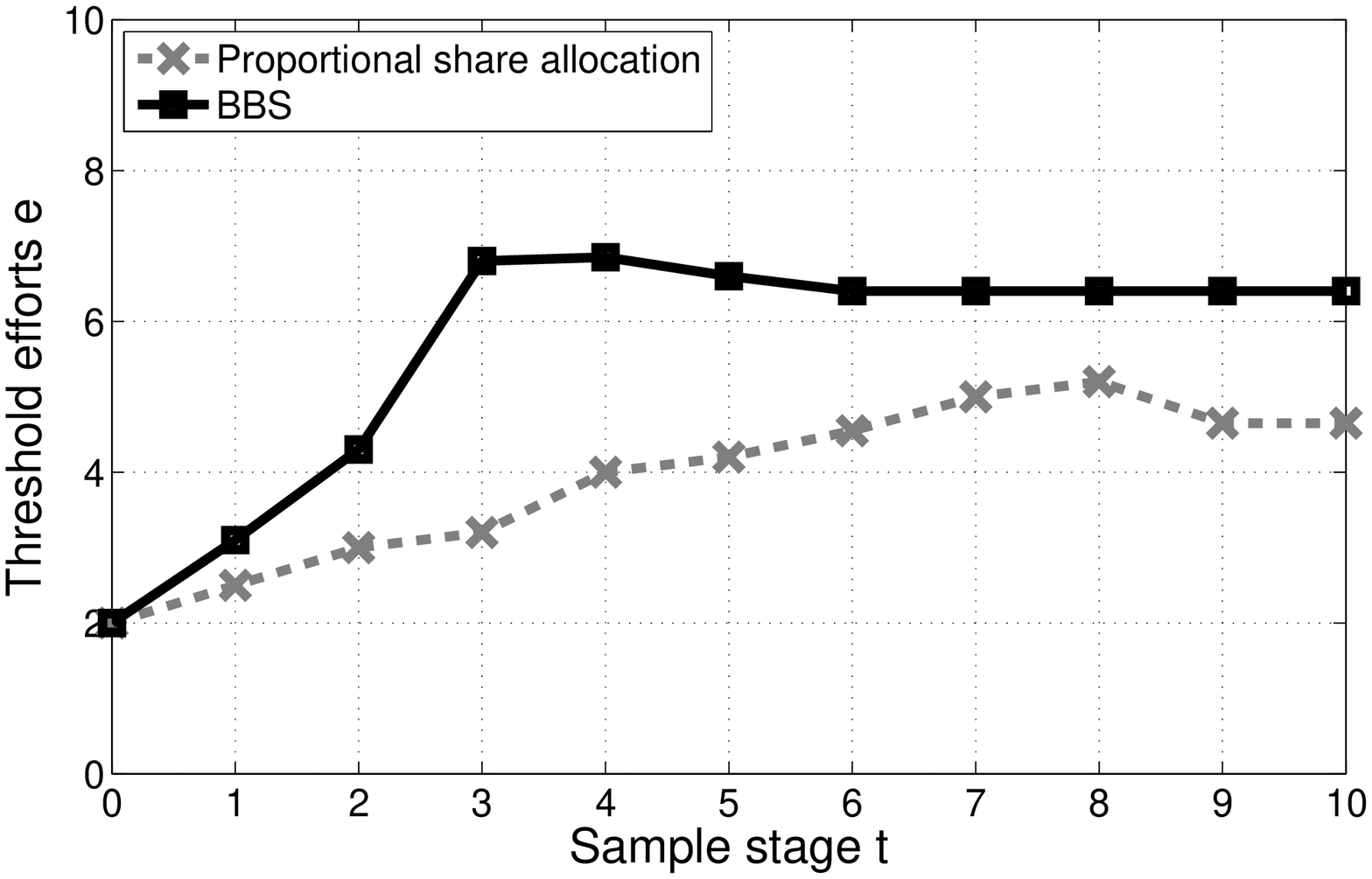}}
\hspace{-0.08in} \subfigure[]{ \label{participants}
\includegraphics[width=1.7in, height=1.3in]{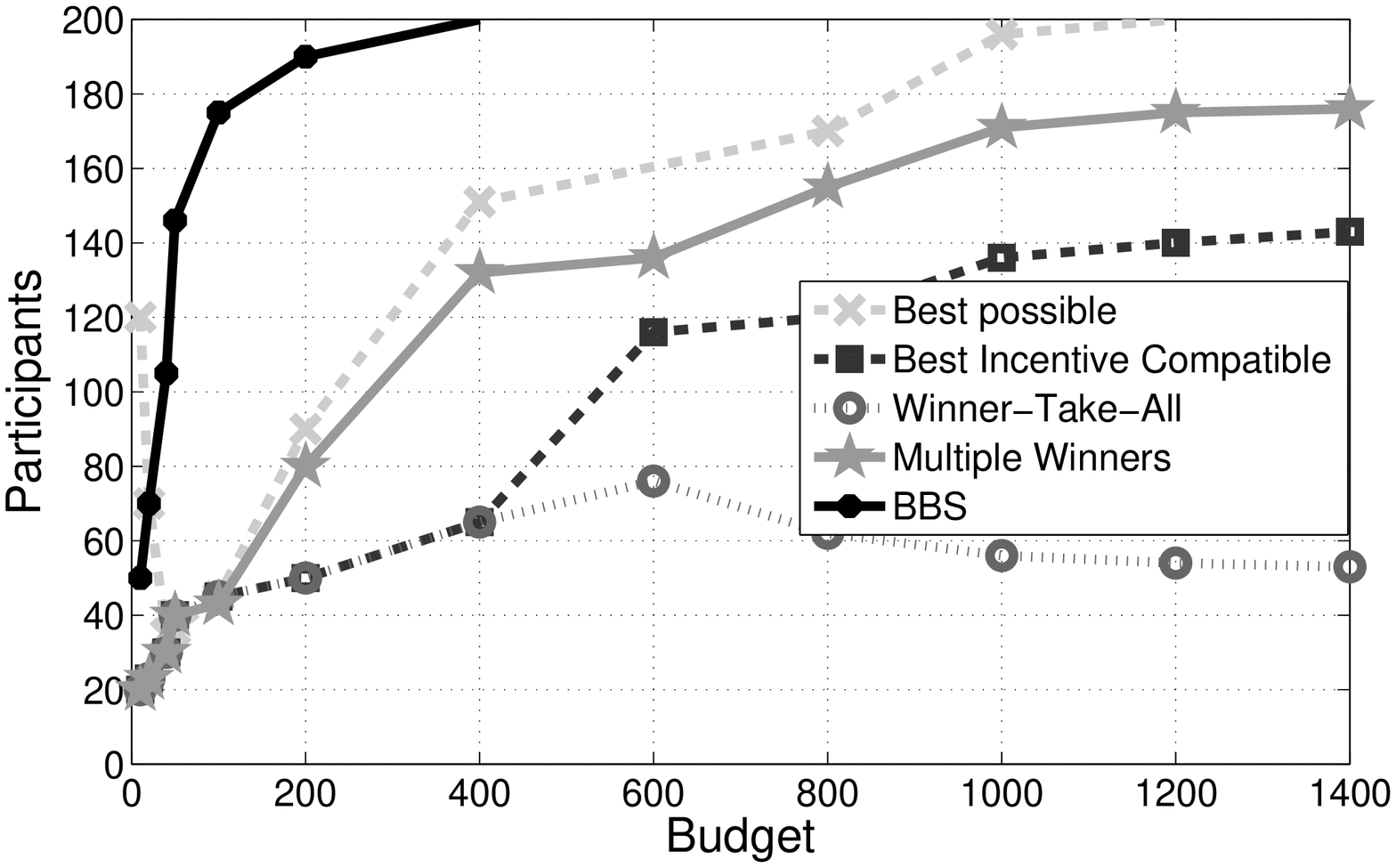}}
\hspace{-0.08in} \subfigure[]{ \label{efforts}
\includegraphics[width=1.7in, height=1.3in]{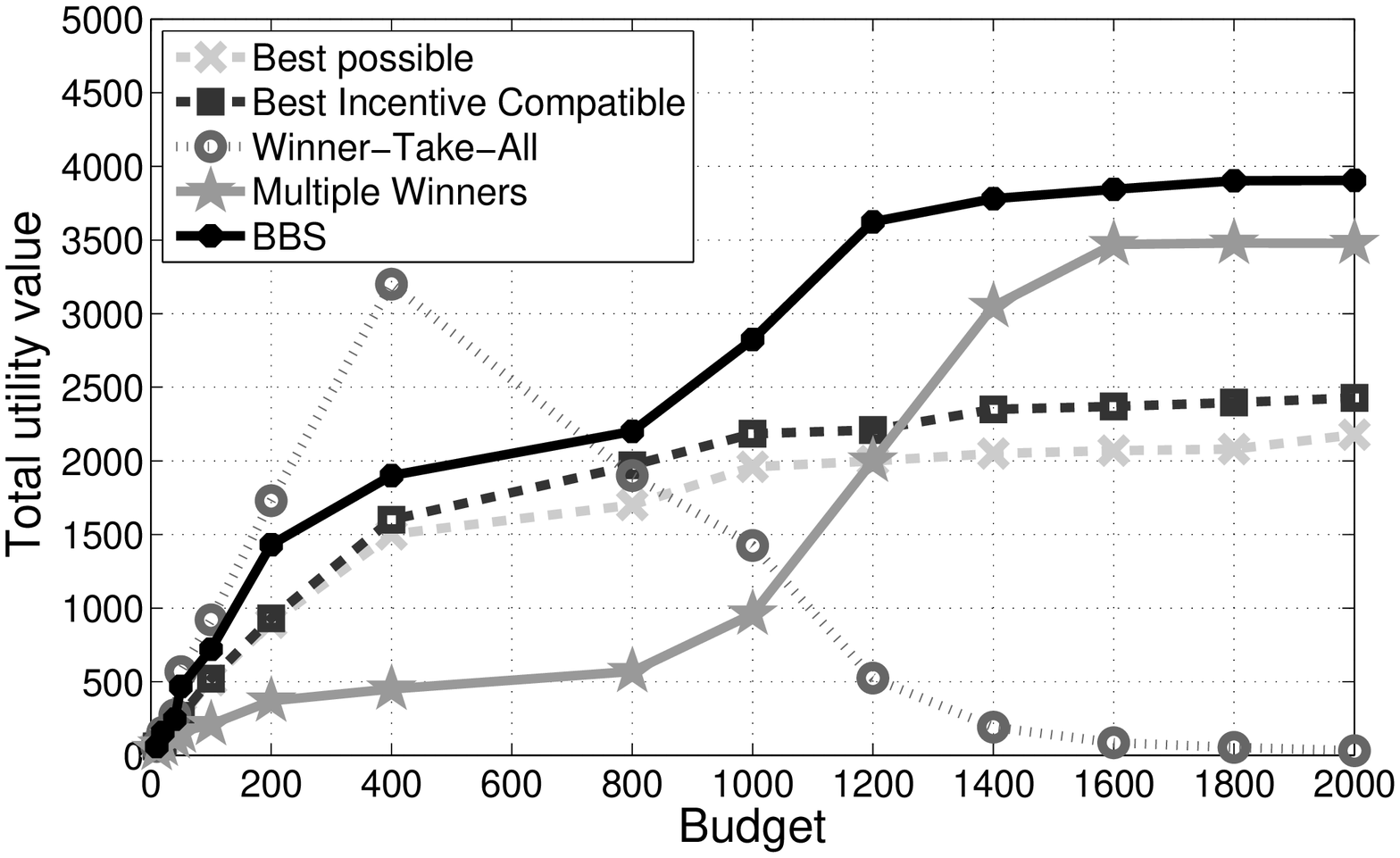}}
\hspace{-0.08in} \subfigure[]{ \label{errors}
\includegraphics[width=1.7in, height=1.3in]{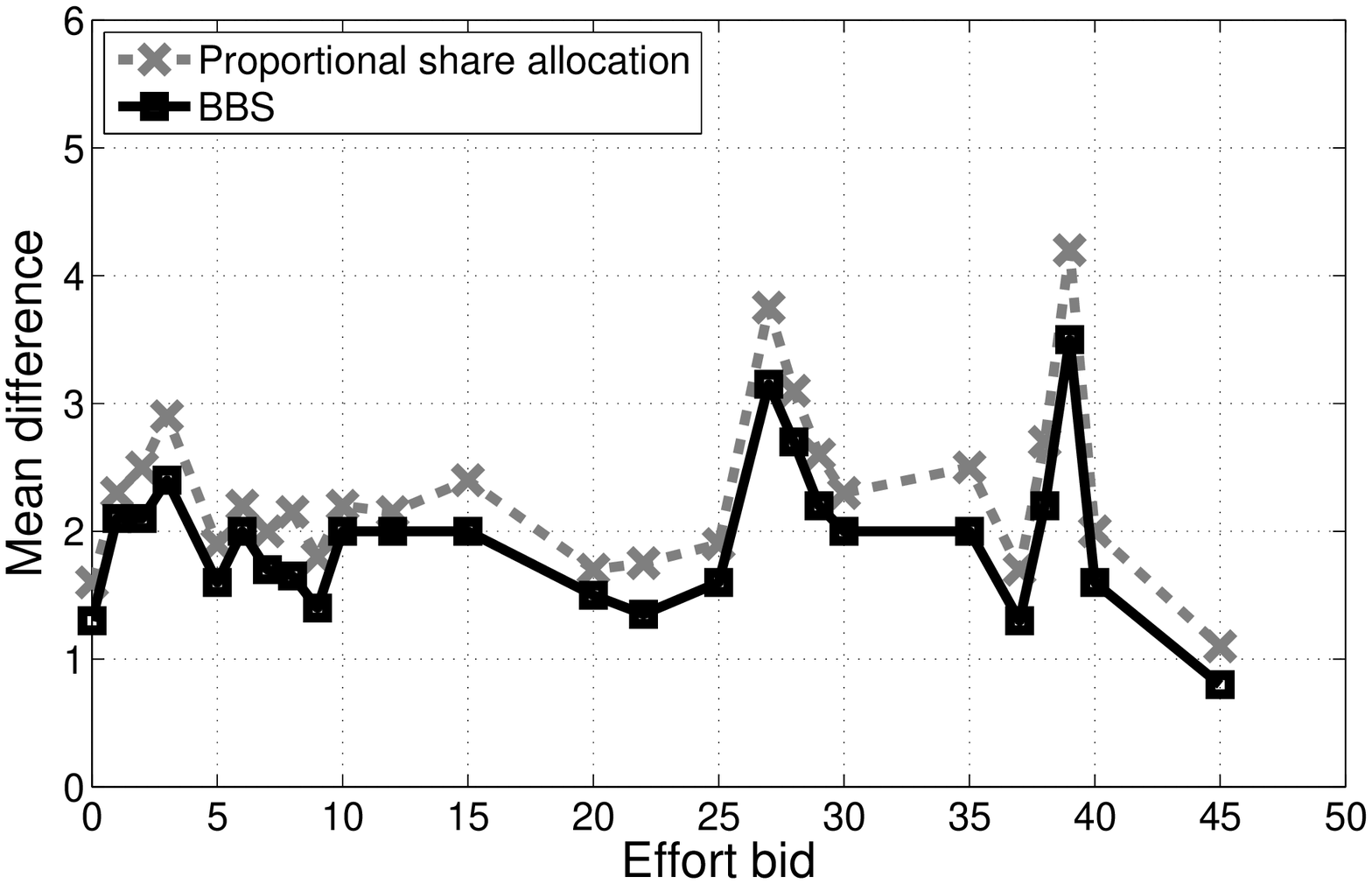}}
\caption{(a)The threshold efforts $e$ versus sample stage $t$;
(b)Effects of user participation; (c)Effects of submission quality;
(d)The difference of submission quality.} \label{fig:random}
\vspace{-15pt}
\end{figure*}

\subsection{Threshold Evolution of Different Mechanisms}
To test the performance of our BBS mechanism, we use the users'
effort bids from all-pay auctions and compare our mechanism against
several benchmarks. Note that in order to show how many users can be
accepted to participate in the crowd sensing given a specified
budget, we only need to computing the values of users' best response
effort bidding function, which is practically obtained according to
the users' arrival sequence. We compare our BBS mechanism against
two benchmarks. One has full knowledge about users' costs, and the
other is the GetEffortThreshold procedure applied offline. We
simulate these algorithms on different budgets to examine the change
in the threshold efforts as the number of users increases in the
sample. In all simulations, we observe that the threshold efforts
converged quickly. Fig. \ref{threshold} shows that the value of the
threshold efforts changes with the stage of the mechanism (the
number of users that submitted their effort bids) on the logarithmic
scale. As we can see, the threshold efforts quickly stabilize and
remain almost constant throughout the running.
%

\subsection{Effects of User Participation}
To examine whether users perform strategic considerations in their
prizing mechanisms, we can observe the distinct difference between
the plots of the different total efforts (bids) in Fig.
\ref{participants} based on different mechanisms. Most of users in
the Winner-Take-All scheme, where a single winner gets all prizes,
drop out of the contest, since the probability of winning the crowd
sensing contest decreased with number of participants. Users in the
Multiple-Winners scheme, where multiple winners get the same prizes,
exert lower effort when there are larger number of participants. In
the reverse auction, bidders accepted in the best possible scheme
increase their bids. (The following drops off since we enforce the
budget constraint). In the best incentive compatible schemes, bids
are lowered, since users bids are rejected. We believe that this is
a strong support for persisting in incentive compatible mechanisms,
since they think that this will increase their profit.
Interestingly, our BBS scheme solves both incentive compatibility
and individual rationality problems.

\subsection{Effects of Submission Quality}
To examine the quality of submission sensing data quality, we plot
the total utility value as a function of different budgets in Fig.
\ref{efforts}. In the Winner-Take-All scheme, most of users drop out
of the contest, since the probability of winning the crowd sensing
contest decreases with the number of participants increasing. Thus,
the total utility value would decreased from its maximum. Users in
the Multiple-Winners scheme, where multiple winners get the same
prizes, exert lower effort and obtain more total utility values when
there are larger numbers of participants as more budgets are
provided. In the reverse auction, bidders accepted in the best
possible scheme increased their bids. In the following, they drop
off since we enforced the budget constraint. In the best incentive
compatible schemes, bids are lowered, since users bids are rejected.
We believe that this is a strong support for persisting in incentive
compatible mechanisms, since they think that this will increase
their total utilities. Interestingly, our BBS scheme solves both IR
and IC problems, therefore, it ensures extensive user participation
and high quality sensing data submission, just as illustrated in
Fig. \ref{participants} and Fig. \ref{efforts}.

Furthermore, we also quantify submission quality by considering
users' mean differences from true on-site data over their effort
bids. Fig. \ref{errors} indicates that our BBS mechanism has a lower
mean errors than the proportional share allocation rule.

\section{Conclusions and Future Work}~\label{Conclude}
In this paper, we present a behavior based incentive strategy to
motivate users to exert the most sensing effort according to their
behavior abilities and willingness for practical crowd sensing
applications. We assess the participants' efforts according to their
context situation (e.g., sensing location and sensing
time). 
We believe that sequential all-pay auctions in our proposed
framework provide a good basis for real crowd sensing applications.
Future research could expand on our BBS mechanism by studying the
effects of privacy protection on participation level and submission
quality. A natural extension of this scheme may be desirable to have
submissions privacy protected and to hide user experience level or
identity.

\section*{Acknowledgment}
This work is supported by the National Natural Science Foundation of
China under Grant No.61332005, No.61272517, No.61133015, the Funds
for Creative Research Groups of China under Grant No.61121001, the
Specialized Research Fund for the Doctoral Program of Higher
Education under Grant No.20120005130002, and the Key Technologies
R\&D Program of China under Grant No.2011BAC12B03.

\bibliographystyle{IEEEtran}

\bibliography{IEEEtran}
\section* {APPDENX}
\noindent\textbf{{Proof of Lemma \ref{expectedsubmodular}:}}\\
From the previous case, we know $U(S)=\sum\limits_{j = 1}^m {\min
\left\{ {1,\sum\nolimits_{i \in S} {s_{i,j} } } \right\}}$. For
$\forall S\subseteq S^{'}\subseteq \mathcal {U}$ and $k\in \mathcal
{U}\setminus S^{'}$, we have
\begin{equation*} 
\begin{split}
U(S\cup\{k\})-U(S)&= \sum\limits_{j = 1}^m {\min \left\{ {\max\{0,
1-\sum_{i \in S} {s_{i,j} }\}, s_{k,j}  }
\right\}} \\
 &\geq\sum\limits_{j = 1}^m {\min \left\{ {\max\{0,
1-\sum_{i \in S^{'}} {s_{i,j} }\}, s_{k,j}  } \right\}}\\
&=U(S^{'}\cup\{k\})-U(S^{'})
 \end{split}
 \end{equation*}
Besides, for $\forall S\subseteq \mathcal {U}$ and $k\in \mathcal
{U}\setminus S$, we have $U(S\cup\{k\})-U(S)\geq 0$. Thus, $U(S)$ is
monotone submodular. Thereby the Lemma \ref{expectedsubmodular}
holds.

\noindent\textbf{{Proof of User $i$ Bidding Function:}}\\
%
In the following, we derive user $i$'s best response effort bidding
function based on the assumption that the behavior ability
distribution function is $F(x)=x^{c}$, where ${0<c<1}$ like
\cite{sela2012multi,liu2011crowdsourcing}. Applying backward
induction, 
we expect that user $n$ will win the auction if the quality of her
solution is higher than or equal to the best quality submission
among all previous submissions, $\max \left\{ {e_j (\theta _j )}
\right\}_{j < n}$, and if her ability is sufficiently high, $\theta
_n  \ge \frac{1}{\overline V}\max \left\{ {e_j (\theta _j )}
\right\}_{j < n}$. If her ability is not high enough, i.e., $\theta
_n < \frac{1}{\overline V}\max \left\{ {e_j (\theta _j )}
\right\}_{j < n}$, then her benefit from winning ($\overline V$) is
less than her bidding cost. In this case, she should bid zero.
Likewise, if he bids zero, indicating that his ability is not high
enough, i.e., $\theta _n < \frac{1}{\overline V}\max \left\{ {e_j
(\theta _j )} \right\}_{j < n}$. On the other hand, user $n-1$ wins
the all-pay auction conditional on her submitting sensing data with
quality at least as high as the best previous submission. Thus,
$\max \left\{ {e_j (\theta _j )} \right\}_{j < n}=e_{n-1}$ holds.
Putting these together, we have: if user $n$ bids zero, his ability
must be not high enough, i.e., $\theta _n < \frac{e_{n-1}}{\overline
V}$. In other words, $F_n (e_n  = 0)=F_n (\theta _n
<\frac{e_{n-1}}{\overline V})=(\frac{e_{n-1}}{\overline V})^{c}$
holds. Likewise, for user $i+1$, we have $F_{i+1} (e_{i+1}  =
0)=(\frac{e_{i}}{\overline V})^{c}$ holds. Applying backward
induction, iteratively for user $n, n-1,\cdots,i+1$, we can obtain
the following equation.
\begin{equation*}
\begin{array}{l}
 \prod\limits_{j = i + 1}^n {F_j (e_j  = 0)}  \\
  = [\frac{{e_i }}{{\overline V}}]^{c(1 - c)^{n - (n - (i + 1))} } [\frac{{e_i }}{{\overline V}}]^{c(1 - c)^{n - (n - (i + 2))} }  \cdots [\frac{{e_i }}{{\overline V}}]^c  \\
  = [\frac{{e_i }}{{\overline V}}]^{1-(1 - c)^{n - i} }  \\
 \end{array}
\end{equation*}
Since the constraint is not binding, the above equation is fed into
equation (\ref{effortbid}) to compute the first-order condition,
thereby, $e_{i}$ is obtained by:
\begin{equation*}
[1 - (1 - c)^{n - i} ][\frac{{e_i }}{{\overline V}}]^{ - (1 - c)^{n
- i} }  - \frac{1}{{\theta _i }} = 0.
\end{equation*}

The second-order condition is obtained by:
 \begin{equation*}
 - \frac{1}{{\overline V}}\{ (1 - c)^{n - i} (1 - (1 - c)^{n - i} )[\frac{{e_i }}{{\overline V}}]^{ - (1 - c)^{n - i} } \}  < 0
 \end{equation*}

Thus, the interior solution is $e_{i}(\theta _i)=\overline V[\theta
_i (1 - (1 - c)^{n - i} )]^{\frac{1}{{(1 - c)^{n - i} }}}$. Let
$d_{i}=(1-c)^{n-i}$. The interior solution is rewritten as
$e_{i}(\theta _i)=\overline V[\theta _i (1 -
d_{i})]^{\frac{1}{d_{i}}}$. Therefore, if we let ${\overleftarrow
{\theta _i } =[(1 - d_{L - th} )\theta _{L - th} ]^{d_i /d_{L - th}
} }$, and ${\overrightarrow {\theta _i }= [(1 - d_{1 - th} )\theta
_{1 - th} ]^{d_i /d_{1 - th} } /c}$, given
$e_{1},e_{2},\cdots,e_{i-1}$, user $i$'s best response effort
bidding function, i.e. the previous expression (\ref{bideffort})
holds evidently.

\end{document}